# X-ray and Neutron Diffraction Investigations of the Structural Phase Transformation Sequence under Electric Field in $0.7Pb(Mg_{1/3}Nb_{2/3})$-$0.3PbTiO_3$ Crystals

Feiming Bai, Naigang Wang, Jiefang Li, and D. Viehland[1], P.M. Gehring[2], Guangyong Xu and G. Shirane[3]

[1]*Dept. of Materials Science and Engineering, Virginia Tech, Blacksburg, VA 24061*
[2]*NIST Center for Neutron Research, NIST, Gaithersburg, MD 20899*
[3]*Department of Physics, Brookhaven National Laboratory, Upton, NY 11973*

The structural phase transformations of $0.7Pb(Mg_{1/3}Nb_{2/3})O_3$-$0.3PbTiO_3$ (PMN-30%PT) have been studied using x-ray diffraction (XRD) and neutron scattering as a function of temperature and electric field. We observe the phase transformational sequence (i) cubic (C)→ tetragonal (T)→ rhombohedral (R) in the zero-field-cooled (or ZFC) condition; (ii) C→T→ monoclinic ($M_C$)→ monoclinic ($M_A$) in the field-cooled (or FC) condition; and (iii) R→$M_A$→$M_C$→T with increasing field at fixed temperature beginning from the ZFC condition. Upon removal of the field, the $M_A$ phase is stable at room temperature in the FC condition, and also in the ZFC condition with increasing field. Several subtleties of our findings are discussed based on results from thermal expansion and dielectric measurements, including (i) the stability of the $M_A$ phase; (ii) a difference in lattice parameters between inside bulk and outside layer regions; and (iii) the diffuse nature of the $M_A$ and $M_C$ phase transition.

## I. Introduction

Single crystals of $Pb(Mg_{1/3}Nb_{2/3})O_3$-$PbTiO_3$ (PMN-PT) and $Pb(Zn_{1/3}Nb_{2/3})O_3$-$PbTiO_3$ (PZN-PT) have attracted much attention as high performance piezoelectric actuator and transducer materials.[1] An electric field induced rhombohedral-to-tetragonal phase transition was first proposed by Park and Shrout to explain the origin of the ultrahigh electromechanical properties. Structural studies of $Pb(Zr_{1-x}Ti_x)O_3$ (PZT) were the first that revealed the existence of a new ferroelectric monoclinic phase, which was sandwiched between the rhombohedral (R) and tetragonal (T) phases near a morphotropic phase boundary (MPB).[2-3]

Two monoclinic phases, $M_A$ and $M_C$, have since been reported in PZN-x%PT.[4-6] The $M_A$ and $M_C$ notation is adopted following Vanderbilt and Cohen.[7] A phase diagram has been reported for PZN-x%PT crystals in the zero-field-cooled (ZFC) condition.[8] Recent neutron diffraction studies of the effect of an electric field (E) on PZN-8%PT by Ohwada et al.[9] have shown that a cubic (C)→T→$M_C$ transformational sequence occurs when field-cooled (FC), and that an R→$M_A$→$M_C$→T sequence takes place with increasing E at 350K beginning from the ZFC condition. An electric field versus temperature (E-T) diagram was constructed based on these experiments.

The same $M_A$ and $M_C$ phases have also been reported in PMN-x%PT[10,11]. Figure 1(a) shows the phase diagram of PMN-x%PT in the ZFC condition, replotted according to recent data published by Noheda et al.[11]. The $M_C$ phase extends from x=31% to x=37%. For x<31%, the phase diagram shows a rhombohedral phase as well as a new phase, designated as X, with an average cubic structure (a=b=c).[12-14] Investigations of poled PMN-35%PT crystals have revealed a $M_A$ phase at room temperature[10].

Diffraction experiments with an in-situ applied electric field[3] together with basic principles calculations[15] on PZT have provided a direct link between the $M_A$ phase and high electromechanical deformations. According to the polarization rotation theory,[16] while the direction of the polarization vector in a conventional ferroelectric tetragonal (or rhombohedral) phase is fixed to the [001] (or [111]) direction, the monoclinic symmetry allows the polarization vector to continuously rotate in a plane. The polarization rotational pathways in the $M_A$ and $M_C$ phases are illustrated in Figure 1(b), where the polarization of the $M_A$ phase is confined to the $(1\bar{1}0)_c$ plane, and the polarization of $M_C$ is confined to the $(010)_c$ plane. Diffraction experiments of PZN-8%PT with an applied in-situ electric field[4,9] have given direct evidence of these polarization rotational pathways and monoclinic phases. However, the transformational sequence has not yet been experimentally established for PMN-x%PT under electric field.

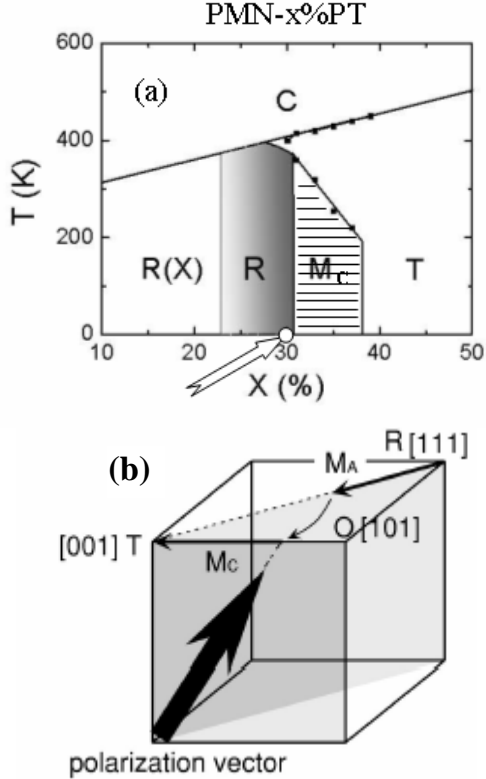

Fig.1. (a) Phase diagram of the PMN-PT solid solution system. The data points come from published results by Noheda et al.[9] The open arrow indicates the concentration studied. (b) Polarization rotation path in the perovskite $M_A$ and $M_C$ unit cell. C, R, T, O and M refer to cubic, rhombohedral, tetragonal, orthorhom-bic and monoclinic regions, respectively.

In this investigation, our focus is on establishing the structural transformation sequence of PMN-30%PT as a function of temperature and electric field. PMN-30%PT, similar to PZN-8%PT, has a composition just outside of the monoclinic phase (see arrow in Figure 1 (a)). Careful experiments have been performed using both x-rays and neutrons, starting from an annealed condition and by (i) cooling the sample from 550K to 300K under constant electric field; and (ii) gradually increasing E at constant temperature. The results unambiguously identify a transformational sequence of C→T→$M_C$→$M_A$ in the FC condition, and of R→$M_A$→$M_C$→T with increasing E at constant temperature in the ZFC condition.

## II. Experimental Details

Neutron and x-ray diffraction measurements were performed on a PMN-30%PT crystal of dimensions 4x4x3 mm$^3$. Crystals were grown by a top-seeded modified Bridgman method, and were obtained from HC Materials (Urbana, IL). All faces of the crystal were polished to a 0.1 μm finish. A gold electrode was then deposited on two 4 x4 mm$^2$ faces by sputtering. The normal to the face on which the electrode was deposited (used to apply an electric field) is designated as (001) or the c axis. Before measurements were begun, the crystal was annealed at 550 K. Careful investigations were performed using both x-rays and neutrons, by starting from this annealed condition in both cases, and by gradually increasing E during sequential FC measurements. The lattice constant of PMN-30%PT in the cubic phase at T=500 K and E=0 kV/cm is $a$=4.024 Å, and thus one reciprocal lattice unit (or 1 r.l.u.) corresponds to $a^*$ (=$b^*$) =$2\pi/a$=1.5616 Å$^{-1}$. All mesh scans presented in this paper are plotted in this reciprocal unit.

The x-ray diffraction (XRD) studies were performed using a Philips MPD high resolution x-ray diffractometer equipped with a two bounce hybrid monochromator, an open 3-circle Eulerian cradle, and a domed hot-stage. A Ge (220) cut crystal was used as analyzer, which had a theta-resolution of ~0.0068° (or 0.43 arcseconds). The x-ray wavelength was that of Cu$_{K\alpha}$ (λ=1.5406 Å) and the x-ray generator was operated at 45 kV and 40 mA. The penetration depth in PMN-30%PT at this x-ray wavelength is on the order of 10 microns. Careful polishing and subsequent annealing were required in order to achieve sharp diffraction peaks – it is important to point this out because prior studies have revealed extremely broad peaks using Cu$_{K\alpha}$ radiation. The neutron scattering experiments were performed on the BT9 triple-axis spectrometer located at the NIST Center for Neutron Research. Measurements were made using a fixed incident neutron energy E$_i$ of 14.7 meV, obtained from the (002) reflection of a PG monochromator, and horizontal beam collimations of 10'-46'-20'-40'. We exploited the (004) reflection of a perfect Ge crystal as analyzer to achieve unusually fine q-resolution near the relaxor (220) Bragg peak, thanks to a nearly perfect matching of the sample and analyzer d-spacings. Close to the (220) Bragg peak, the q–resolution along the wave vector direction is about 0.0012Å$^{-1}$ ($\Delta q/q \approx 2 \times 10^{-4}$).[12] Both x-ray and neutron measurements were performed as a function of temperature and dc electrical bias. Extremely sharp q-resolution is needed to detect the subtle broadening and splitting of the Bragg peaks using either x-ray or neutron probes.

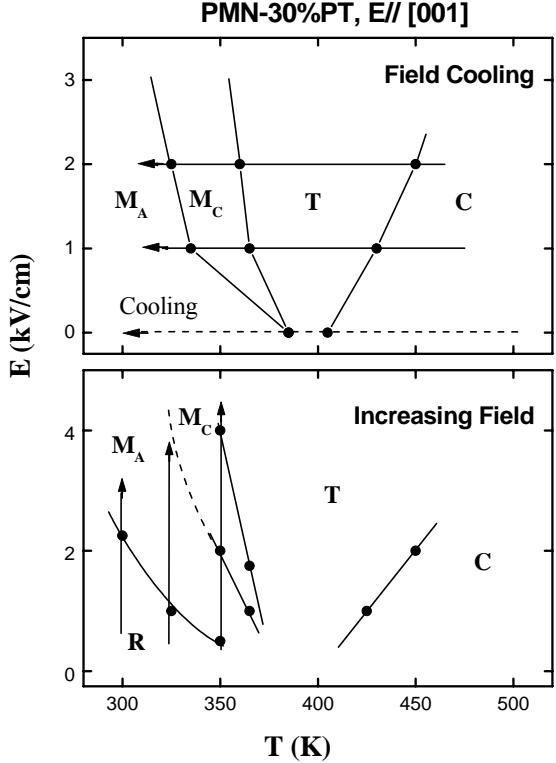
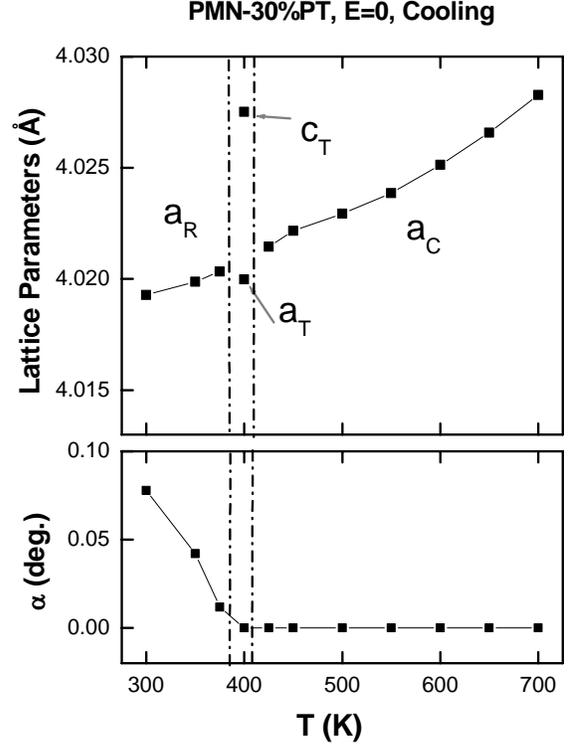

Fig.2. *E-T* diagram. Top panel is obtained from FC structural measurements. Bottom panel shows data from the increasing electric-field process after ZFC. Arrows indicate the scanning directions and ranges of the corresponding measurement sequences. Circles represent the transition temperatures and fields determined from each sequence.

FIG.3. The dependence of the lattice parameters (top panel) and α (bottom panel) on temperature under zero electric field.

## III. Identification of Phase Transformational Sequence

Our electric field-temperature measurements are summarized in Figure 2. This is done for convenience of the readers – raw data will be presented in the following subsections. The top panel of this figure gives the field-cooled diagram, where measurements were made under a constant field on cooling from 500 K, whereas the bottom panel was obtained by increasing E beginning from the ZFC condition at each fixed temperature. Circles represent the transition temperatures and fields determined from each increasing field sequence. Arrows are used to indicate the scanning direction and range of the corresponding measurement sequence.

### III.1 XRD Investigations

*Phase stability in zero-field-cooled condition*

The temperature dependence of the lattice parameter was investigated under zero electric field (E=0kV/cm). The specimen was first heated up to 700 K, and it was confirmed that the structure was cubic. Measurements were then made on cooling. A cubic to tetragonal phase transition was observed near 405 K associated with 90° domain formation, which was confirmed by observing a peak splitting of the (200) reflection. By fitting the (200) reflection with a double Gaussian function, we obtained the temperature dependence of the lattice constants $c_T$ and $a_T$, as shown in the top panel of Fig. 3. On further cooling, a subsequent tetragonal to rhombohedral transformation was found near 385 K. This was manifested by the development of a splitting of the (220) reflection, and a simultaneous disappearance of the (200) peak splitting. The rhombohedral lattice parameters and tilt angle (α) were calculated by fitting the (220) reflection to (220) and ($2\bar{2}0$) peaks. The temperature dependence of α is shown in the bottom panel of Figure 3. Our x-ray results under zero field are in close agreement with the x-ray powder diffraction results on PMN-30%PT of Noheda *et al.*[11]

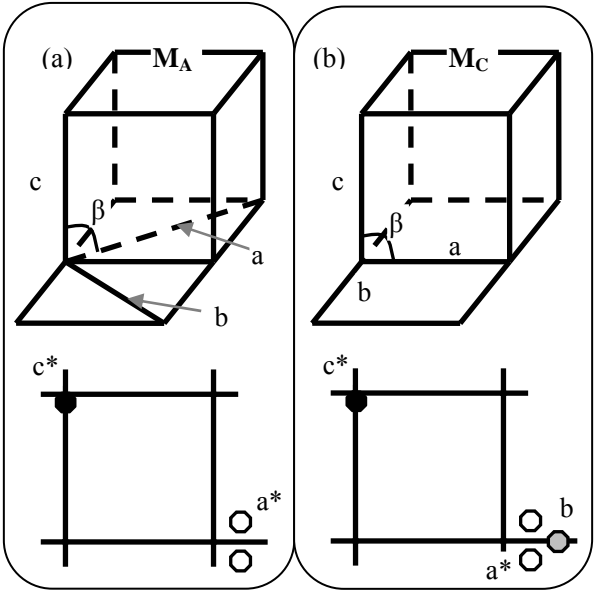

Fig.4. Sketch of the unit cell and domain configuration in the reciprocal (h 0 l) plane for monoclinic phases, (a) top: unit cell of $M_A$ phase; bottom: domain configuration in reciprocal space, illustrating the two *a* domains of $M_A$; and (b) top: unit cell of $M_C$ phase; bottom: domain configuration in reciprocal space, illustrating the two *a* domains (unshaded) and one *b* domain (shaded) of $M_C$ phase.

*Phase stability in field-cooled condition*

The temperature dependence of the lattice parameter was investigated under electric fields of 1 and 2 kV/cm. The specimen was first heated up to 550 K, where it was confirmed that the structure was cubic. An electric field was then applied and measurements were made on cooling. For E=1 kV/cm, a cubic to tetragonal phase transition was observed on cooling near 430 K, and a tetragonal to monoclinic $M_C$ transformation was found near 365 K. On further cooling a subsequent monoclinic $M_C$ to monoclinic $M_A$ transition occurred. After increasing E to 2kV/cm, the T→$M_C$ and $M_A$→$M_C$ transition temperatures decreased, and the phase stability ranges of both T and $M_C$ phases increased.

A sketch of the unit cells and domain configurations in the reciprocal (h0l) plane for the $M_A$ and $M_C$ phases is shown in Figures 4(a) and (b), respectively. For the $M_A$ phase, $a_m$ and $b_m$ lie along the pseudo-cubic $[\bar{1}\bar{1}0]$ and $[1\bar{1}0]$ directions, and the unit cell is doubled in volume with respect to the pseudo-cubic unit cell. For the $M_C$ phase, $a_m$ and $b_m$ lie along the [100] and [010] directions, and the unit cell is primitive. In both cases, the angle between $a_m$ and $c_m$ is defined as β. Usually, monoclinic symmetry leads to a very complicated domain configuration. However, once the field is applied, the *c* axis is fixed along the field direction. The monoclinic domain configuration then consists of two *b* domains related by a 90° rotation around the *c* axis, and each of the *b* domains contains two *a* domains in which $a_m$ forms angles of either β or 180°- β. In the (H0L)$_{cubic}$ zone, only two *a* domains can be observed for the $M_A$ phase, as shown in Figure 4(a); whereas one *b* domain and two *a* domains can be observed for the $M_C$ phase, as shown in Figure 4(b).

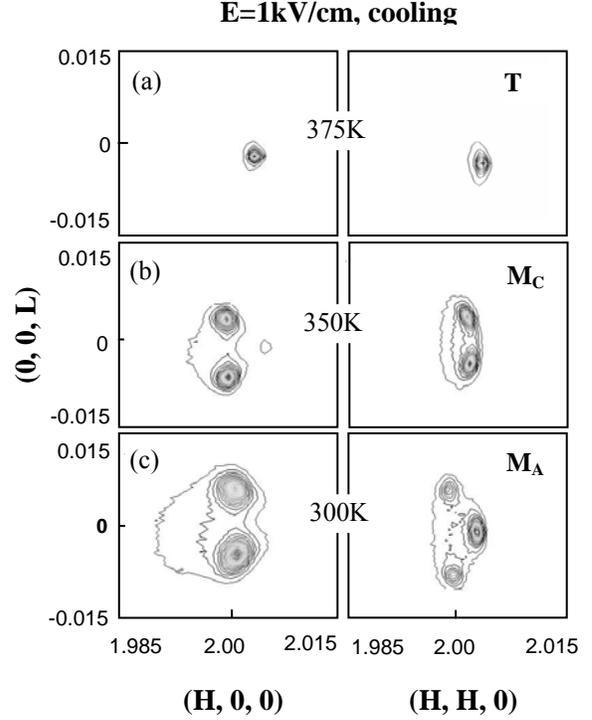

Fig.5. Mesh scans around the (200) and (220) reciprocal lattice positions at different temperatures in field-cooled process.

To best illustrate the observed transformational sequence, XRD mesh scans around (200) and (220) reflections are shown in Figure 5 taken at temperatures of 375, 350 and 300 K. These scans were all obtained under an applied dc electrical bias of E=1 kV/cm. For T=375 K, the lattice constant $c_T$ is elongated, whereas $a_T$ is contracted. This indicates a phase with tetragonal symmetry. For T=350 K, the (200) reflection was found to split into three peaks, consisting of two (200) peaks and a single (020) peak; whereas, the (220) reflection was found to be split into two peaks. These results indicate a phase with monoclinic $M_C$ symmetry. On further cooling, significant changes in the mesh scans were found. For T=300 K, the (200) reflection was found to split only into two peaks, which can be attributed to the presence of two domains, whereas the (220) reflection was found to split into three peaks. This indicates a phase with monoclinic $M_A$ symmetry. The room temperature mesh scans are consistent with those previously reported by Ye et al.[10] for PMN-35%PT crystals, demonstrating that the $M_A$ phase is stable in the FC condition. Moreover, our results also give conclusive and direct evidence of an $M_C \rightarrow M_A$ transition on cooling in the FC condition for PMN-30%PT. This is different from the results for PZN-8%PT single crystals, where an $M_C \rightarrow M_A$ transition was not observed in the FC condition.[9]

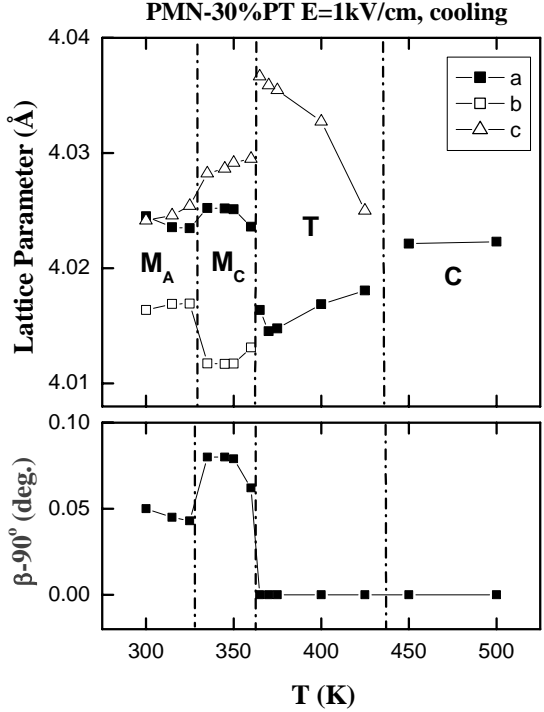

Fig.6. Temperature dependence of the lattice parameters (top panel), and 90°-β (bottom panel) observed in field-cooled process. For the $M_A$ phase, the lattice parameters $a_{Ma}/\sqrt{2}$, $b_{Ma}/\sqrt{2}$ and $c_{Ma}$ are plotted; whereas, for the $M_C$ phase the lattice parameters $a_{Mc}$, $b_{Mc}$ and $c_{Mc}$ are plotted. Solid lines drawn through the data points are guides to the eyes.

Figure 6 shows the temperature dependence of the structural data in the FC condition for E=1 kV/cm. The top panel of this figure shows the lattice parameters, and the monoclinic tilt angle (β−90°) is shown in the bottom panel. The lattice constant $c_T$ ($a_T$) gradually increases (decreases) with decreasing temperature. Near T=365 K, where the T→$M_C$ transition occurs, the value of the lattice constants abruptly changes and a monoclinic tilt angle of $β_{Mc}$−90° ≈0.08° forms between the (001) and (100). In the $M_C$ phase region, the lattice parameters $a_{Mc}$, $b_{Mc}$, $c_{Mc}$, and $β_{Mc}$ are relatively temperature independent over the range of temperatures investigated. The value of $b_{Mc}$ can be viewed as a natural extension of the $a_T$ lattice parameter; the value of $a_{Mc}$ is close to the value of the cubic lattice parameter, whereas the value of $c_{Mc}$ is notably different than either $c_T$, $a_T$, or $a_c$. Near T=330 K, the lattice constants and the tilt angle abruptly change, where the $M_C \rightarrow M_A$ transition occurs. Again, in the $M_A$ phase region, it was found that lattice parameters are only weakly temperature dependent over the range of temperatures investigated.

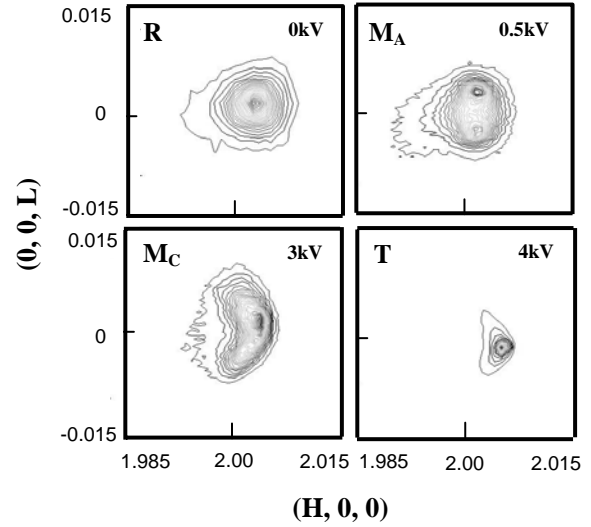

Fig.7. (200) mesh scan at 350K with increasing field, which clearly shows a sequential phase transition from R→$M_A$→$M_C$→T.

*Phase Stability at fixed temperatures with increasing E*

The field dependence of the lattice structure was investigated at various temperatures. The specimen was first heated up to 550 K and then cooled under zero field. This was done at the beginning of measurements at each temperature, thus ensuring that the specimen was always properly zero-field cooled. XRD mesh scans were then performed at various dc electric biases for 0≤E≤4 kV/cm. Both R→$M_A$ and $M_A$→$M_C$ transitions were observed with increasing E.

TABLE 1. Lattice parameter for the PMN-30%PT at 350K with increasing electric filed, measured by XRD. Errors = ±0.002Å.

|  | a (Å) | b (Å) | α(=γ) (°) | β (°) | Phase |
|---|---|---|---|---|---|
| ZFC from 550K, E=0 | 4.020 |  |  |  | R |
| E=0.5kV/cm | 4.023 |  | 90 | 90.08 | $M_A$ |
| E=2kV/cm | 4.019 | 4.014 | 90 | 90.09 | $M_C$ |
| E=4kV/cm | 4.015 | 4.015 | 90 | 90 | T |
| After removed, E=0 |  |  |  |  | $M_A$ |

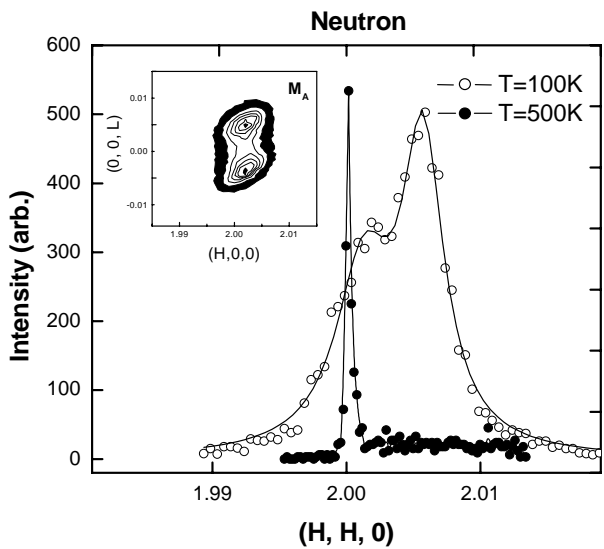

Fig.8. Neutron (220) profiles for PMN-30%PT. The sample was cooled under E=0. The solid lines are fits described in the text. The inset shows a neutron intensity contour around the pseudo-cubic (200) reflection in the H0L zone at 300K, which confirms the existence of the $M_A$ phase in the FC condition. These data were all taken using a perfect crystal Ge (004) analyzer[17].

Mesh scans around (200) are shown in Fig.7 taken at 0, 0.5, 3 and 4kV/cm. These scans were all obtained at T=350 K. The corresponding lattice parameters were listed in Table 1. For E=0 kV/cm, a rhombohedral phase was found. Under E=0.5 kV/cm, the (200) reflection was found to be split into two peaks. This indicates an R→$M_A$ transition with increasing E. Under E=3 kV/cm, the (200) was found to be split into three peaks, revealing a monoclinic $M_C$ phase. And under E=4 kV/cm, the (200) was found to become to one peak, revealing a tetragonal T phase.

These changes in the mesh scans provide conclusive evidence of an R→$M_A$→$M_C$→T phase transition sequence with increasing E starting from the ZFC condition. The field at which the R→$M_A$, $M_A$→$M_C$ and $M_C$→T transitions occur varies with temperature, and is summarized in the E-T diagram of Figure 2(b). Upon removal of the field, the R phase does not recover, but instead a monoclinic phase appears. For T<350 K, the $M_A$ phase is recovered; whereas for T>350 K, the $M_C$ phase is recovered. Thus, the $M_A$ phase dominates the E-T diagram. This is different from the previous results for PZN-8%PT[9], where the $M_C$ phase recovers after removal of the field, instead of the $M_A$ phase.

## III.2 Neutron Investigations

Neutron scattering investigations were also performed in the zero-field-cooled state. The crystal was first heated up to 600 K, where again it was confirmed that the structure was cubic. Measurements of both (220) and (002) reflections were then made on cooling between 600K and 50K. A cubic to tetragonal phase transition is observed near 410 K, and confirmed by the peak splitting of the (002) reflection, similar to that found by XRD. A tetragonal to rhombohedral phase transition is observed near 385 K, as evidenced by the disappearance of the (002) splitting, and the development of a (220) splitting. The rhombohedral phase persists down to 50 K.

Figure 8 shows (220) scans taken at 500K and 100K. The single peak at 500K confirms that the structure is cubic, while the splitting of the (220) is clearly evident in the low temperature rhombohedral phase. In addition, a dramatic change in the line width was found on cooling. The line width was nearly an order of magnitude broader in the rhombohedral phase, relative to that in the cubic phase.

By fitting the (002) reflection with a double Gaussian function, we obtained the tetragonal lattice constants $c_T$ and $a_T$. The lattice parameter and tilt angle ($\alpha$) of the rhombohedral phase were calculated by fitting the (220) reflection to (220) and ($2\bar{2}0$) peaks. A summary of the temperature dependence of the lattice parameters is presented in Table 2. The neutron results clearly show that the rhombohedral phase is stable at room temperature, in agreement with our XRD studies and previous investigations by Noheda et al.[11]

Mesh scans of a poled PMN-30%PT single crystal were also obtained by neutron scattering. A (200) peak splitting along the transverse direction was found, as can be seen in the inset of Figure 8. The (200) splitting can be attributed to two a domains. This neutron mesh scan is nearly identical to the XRD mesh scan shown in Figure 6. The neutron mesh scan shows that the $M_A$ phase is stable in the bulk of the crystal.

TABLE 2. Lattice parameters of PMN-30%PT under zero-field, measured by neutron scattering. Errors= ±0.001Å.

| T (K) | Phase | a (Å) | b (Å) | c (Å) | α(=γ) (°) | β (°) |
|---|---|---|---|---|---|---|
| 600 | C | 4.022 | 4.022 | 4.022 | 90 | 90 |
| 550 | C | 4.020 | 4.020 | 4.020 | 90 | 90 |
| 450 | C | 4.019 | 4.019 | 4.019 | 90 | 90 |
| 400 | T | 4.017 | 4.017 | 4.023 | 90 | 90 |
| 350 | R | 4.019 | 4.019 | 4.019 | 89.96 | 89.96 |
| 300 | R | 4.019 | 4.019 | 4.019 | 89.91 | 89.91 |
| 200 | R | 4.018 | 4.018 | 4.018 | 89.87 | 89.87 |
| 100 | R | 4.017 | 4.017 | 4.017 | 89.84 | 89.84 |

## IV. Discussion

Our results for PMN-30%PT in the ZFC condition demonstrate a phase transformational sequence of C→T→R with decreasing temperature ($T_c \approx 405K$). The results are in agreement with the phase diagram given in Figure 1(a), which provides a confirmation for the composition of the crystal. This is important to know for sure, as if x had been slightly higher, $M_C$ could have been the stable ground state, rather than R.

### *The stability of the $M_A$ phase*

With decreasing temperature under a constant applied field (i.e., in the FC condition), we find that PMN-30%PT undergoes the phase transformational sequence C→T→$M_C$→$M_A$. This sequence is similar to that observed in PZN-8%PT[9], except that our x-ray and neutron investigations reveal an $M_C$→$M_A$ transition with decreasing temperature, which was not found for PZN-8%PT. Moreover, with increasing field at fixed temperature starting from the ZFC condition, we find that PMN-30%PT exhibits the phase transformational sequence R→$M_A$→$M_C$→T. The R→$M_A$ transition is irreversible upon removal of the field, whereas the $M_A$→$M_C$ transition is reversible. Similar studies of PMN-25%PT and PMN-35%PT (data not shown) also indicate that the $M_A$ phase is stable at room temperature in the after-poled condition. Clearly, compared with the MPB compositions of PZN-x%PT, the $M_A$ phase dominates the *E-T* diagram of PMN-x%PT, and not the $M_C$ phase. Ohwada *et al.*[9] explained the $M_C$ phase in PZN-8%PT by a hidden orthorhombic symmetry, located on the T-$M_C$-O polarization rotational pathway (see Figure 1(b)). However, more thought is required to explain the reversibility of the $M_C$→$M_A$ transition in PMN-x%PT after removal of the field, since the $M_C$ phase is currently believed to be the ground state at the MPB.

Recent investigations by Singh *et al.*[18] on polycrystalline PMN-x%PT for $0.27 \leq x \leq 0.30$ have reported a monoclinic $M_B$ phase at room temperature. In the $M_B$ phase, the polarization would be constrained to the (100)$_c$ plane, and there would be two *a* domains in the reciprocal (hk0) plane where the lattice parameters fulfill $a_{Mb} > c_{Mb}$. Our investigations have failed to find an $M_B$ phase, either in the ZFC condition (where we found the R phase) or in the FC condition (where we found two *a* domains with $a_{Mb} < c_{Mb}$, i.e., the $M_A$ phase). However, recent investigations by Viehland and Li[19] have indicated that the $M_B$ phase can be induced *only* by an electric field applied along (011)$_c$, where the $M_A$ phase is recovered on removal of E.

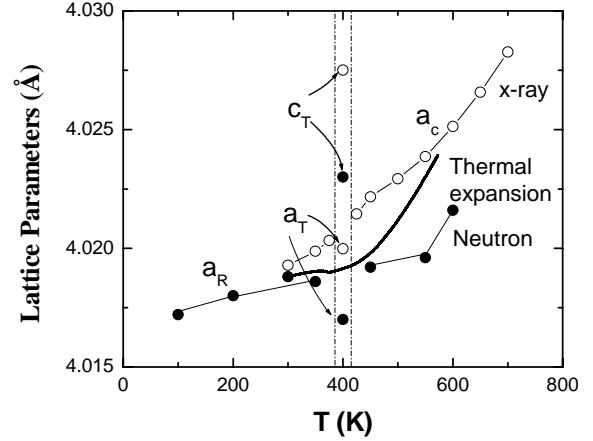

Fig.9. Lattice parameters of PMN-30%PT as a function of temperature measured by XRD, thermal expansion, and neutron scattering.

### *Difference between inside layer and outside layer*

In the ZFC condition, the lattice structure of PMN-30%PT, measured using both x-ray and neutron probes, are similar. Temperature dependent x-ray and neutron measurements revealed the same C→T→R transformation sequence. However, the lattice parameters were found to be different in the x-ray and neutron data. This is illustrated in Figure 9. Those measured using x-rays were larger than those by neutrons. The effect was most pronounced in the tetragonal and cubic phases. On cooling in the rhombohedral phase, the lattice constants for both measurement methods became nearly equivalent. The evolution of the lattice strain at high temperatures was also measured using thermal expansion, as also shown in Figure 9. The lattice parameters were estimated from the thermal expansion as 4.019(1+ε), where ε is the thermal expansion strain and 4.019 Å is the value of the rhombohedral lattice constant $a_R$ at room temperature as determined by neutrons. For T<425 K, the lattice parameter estimated from the thermal expansion is nearly equivalent to that determined by neutrons. However, at higher temperatures, the values determined by thermal expansion approach those for x-rays, both being notably larger than those for neutrons. Linearity in the lattice strain was evident in both the x-ray and thermal expansion data at quite high temperatures, i.e., for $T \geq T_{Burns}$ where $T_{Burns}$ is the temperature of the onset of local polarization.[20]

For PZN-x%PT, previous investigations have surprisingly revealed different structures for the "outside layer" and "inside bulk" regions.[5,21,22] This structural variability was attributed to differences in the volume of the specimen probed. Neutrons have a large spot size and penetrate the entire specimen, thus their signal is representative of the volume; whereas x-rays have a small spot size and a shallower penetration depth, thus their signal is representative of the region close to the crystal surface.

The difference between the lattice parameters of inside bulk and outside layer regions in Fig. 9 can be explained by a difference in stress state. In the outside layer region, internal stresses can be relaxed. This will result in an expansion of the lattice parameter of the outside layer regions, relative to that of the inside bulk.

*Diffuse nature of phase transformational sequence*

The *E-T* diagrams in Figure 2 show that the temperature stability range of the $M_A$, $M_C$, and T phases is significantly increased by the application of an electric field. This can be further noted by comparisons of the lattice parameter data for the ZFC state in Figure 3 with that for the FC state (E=1 kV/cm) in Figure 5.

We performed dielectric constant measurements under the same temperature and field conditions as that for the ZFC and FC lattice parameter data. Figure 10 shows the dielectric constant as a function of temperature for (a) E=0 kV/cm on cooling, and (b) E=1 kV/cm on cooling. The dielectric response in both cases is a single, broad featureless peak, which is typical of a diffuse phase transition. The temperature of the dielectric constant maximum ($T_{max}$) did *not* change between the ZFC and FC measurements – $T_{max}$ is ~405K in both cases. In this regard, the dielectric constant data reveal a notable difference with respect to the corresponding lattice parameters. Previous studies of PZN-8%PT by Ohwada *et al.*[9] have shown a similar large change of the C→T boundary with increasing E in the *E-T* diagrams; in addition, recent dielectric investigations of PMN-30%PT have shown a field independence of $T_{max}$, similar to that shown in Figure 10.[23] However, recent dielectric studies of PMN-x%PT for 29<x<35 have shown a shift in the MPB due to application of E along the (001), and an increase in $T_{max}$.[24]

One would expect that the changes in the lattice parameters would occur at temperatures below $T_{max}$. This is required either for a 1st or 2nd order transition. However, in the FC state, our data show that a tetragonal splitting develops at T>$T_{max}$. In diffuse transitions, local polar regions are believed to have transition temperatures notably greater than that of the dielectric maximum.[25] According to the theory of diffuse transitions, the prototypic cubic and low temperature ferroelectric phases coexist over a broad temperature range, where the relative volume fractions of the coexisting phases change with temperature. However, our structural investigations demonstrate a single phase tetragonal region over the temperature range of 365<T<430K, even though the observed $T_{max}$ from the dielectric constant was ~405K. Recently, an alternative theory of structurally mixed states has been suggested, which is inhomogeneous on the nano-scale, but homogeneous on the macro-scale.[26] According to this theory, changes in lattice parameters can be gradual and occur over a broad temperature range.

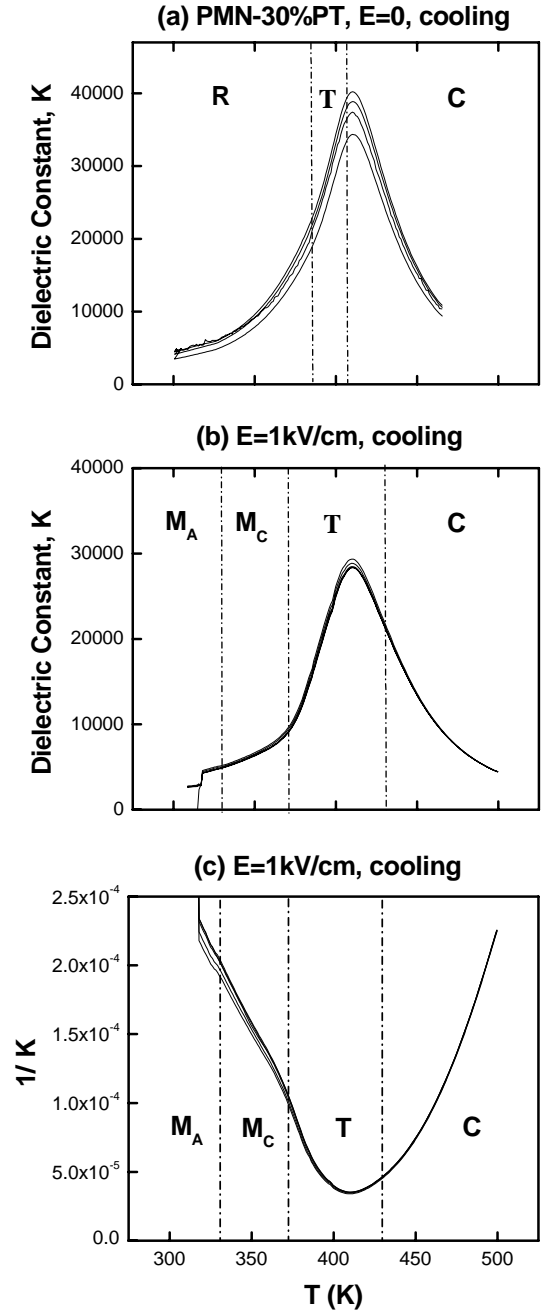

Fig.10. Temperature dependence of the dielectric constant (a) in zero field cooled process; (b) in field cooled process; and (c) the Curie-Weiss plot for field cooled process.

Indications of non-conventionality were also found in the $M_A \rightarrow M_C$ transition. It is important to first remind ourselves that the $M_C$ and $M_A$ phases do not have subgroup relationships between the symmetries of their point groups. Glazer has recently noted the relevance of this fact to MPB transitions in ferroelectrics.[27] Thus, in a displacive transformation, originating from an m3m prototypic symmetry, the $M_C \rightarrow M_A$ transition *is required* to be first order. A common way to determine the order of a ferroelectric transition is by a Curie-Weiss plot of the dielectric constant (K). Figure 10 (c) shows a Curie-Weiss plot, $K^{-1}$ vs. T. Neither, a 1$^{st}$ or 2$^{nd}$ order transition was found at the $M_C \rightarrow M_A$ transition. Rather, the data is featureless, indicative of a diffuse $M_C \rightarrow M_A$ transition and that this transition can not be understood as a conventional displacive structural phase transition.

In conclusion, we have established that the phase transformational sequence of PMN-30%PT is (i) $C \rightarrow T \rightarrow M_C \rightarrow M_A$ on cooling in the FC condition; and (ii) $R \rightarrow M_A \rightarrow M_C \rightarrow T$ with increasing E at constant temperature starting from a ZFC condition. The $R \rightarrow M_A$ transition is irreversible, but the $M_A \rightarrow M_C$ transition is reversible upon removal of the electric field. As a consequence $M_A$ is the dominant phase for $T<T_c$ in the *E-T* diagram. Evidence is also given that the phase transformation sequence of PMN-30%PT is diffusive.

**Acknowledgements**

We would like to thank M. Glazer, L.E. Cross, and A.G. Khachaturyan for stimulating discussions, and H.C. Materials for providing the single crystals. Financial support from the U.S. Department of Energy under contract No. DE-AC02-98CH10886 and Office of Naval Research under grants N000140210340, N000140210126, and MURI N000140110761 is also gratefully acknowledged. We also acknowledge the NIST Center for Neutron Research for providing the neutron scattering facilities used in this study.